\documentclass[12pt]{article}
\usepackage{amsmath,amssymb}
\usepackage{geometry}
\usepackage{graphicx}
\usepackage{setspace}
\usepackage{hyperref}
\usepackage{color}
\usepackage{cite}

\begin{document}

\title{Discussion summary: \\
AI and the Research--Education Environment of Physics}
\date{December 11, 2025\\
Kavli Institute for Theoretical Physics, Santa Barbara\\
Based on discussions in the program\\
``Generative AI for High and Low Energy Physics''}
\maketitle

\begin{abstract}
    In the current era of AI transforming the research-education environment of physics, variety of issues and concerns arise. The KITP program ``Generative AI for High and Low Energy Physics'' offered a discussion session on this, and here presented is a summary of the opinions provided in the discussion. The material is formulated such that it can serve as a starting point for further discussions in readers' research community / institution / group.
\end{abstract}

\begin{figure}[h]
    \centering
    \includegraphics[width=0.95\textwidth]{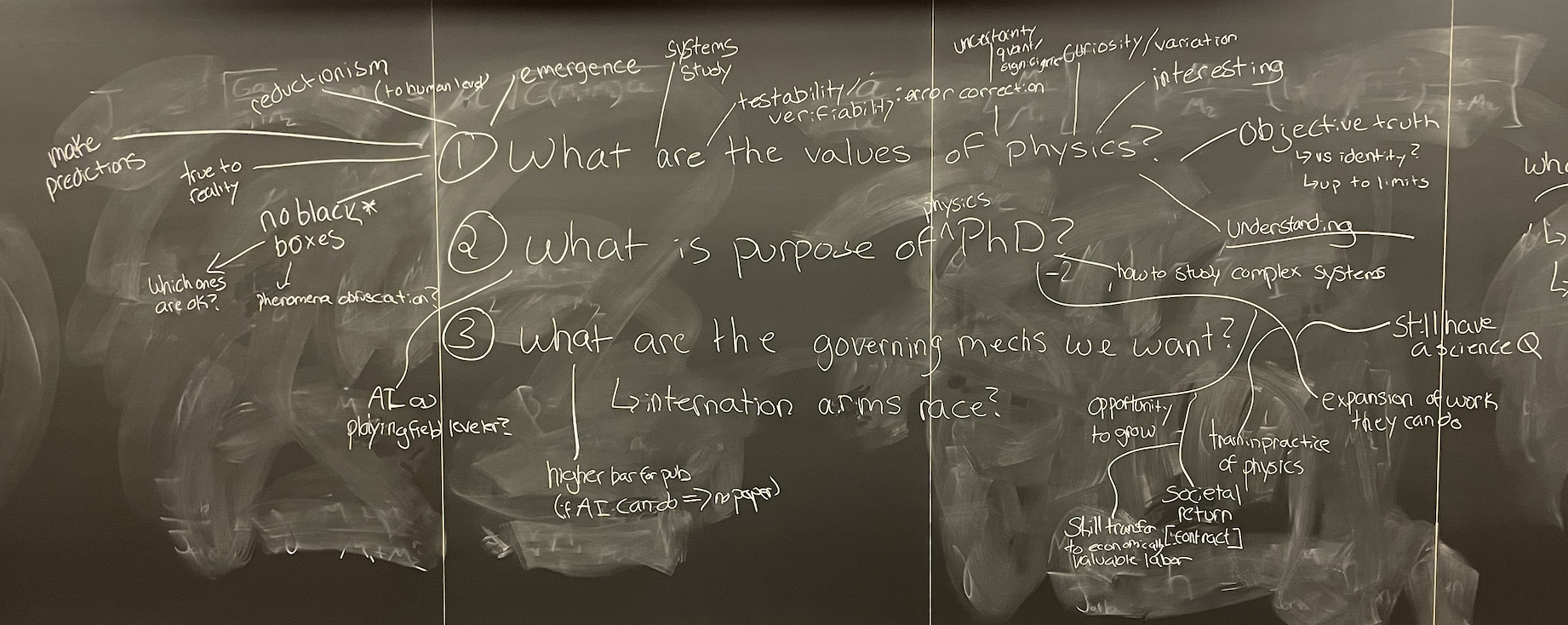}
    \caption{Photograph of the discussion board from the session led by Savannah Thais (Hunter College), summarizing the framing questions.}
\end{figure}

\clearpage

\noindent
List of the participants who agreed to be listed: 

Savannah Thais (Hunter College) (Discussion lead)

Koji Hashimoto (Kyoto University) (1st draft)

David S. Berman (Queen Mary University of London)

Estelle Inack (Perimeter Institute, yiyaniQ \& University of Waterloo)

Jessica N. Howard (UC Santa Barbara / Kavli Institute for Theoretical Physics)

Gregor Kasieczka (Universit\"at Hamburg)

Anindita Maiti (Perimeter Institute)

Garrett W. Merz (University of Wisconsin-Madison)
    
Toledo, Javier (TRIUMF \& Perimeter Institute)

\noindent
More than 20 participants, from different continents, joined the discussion.

\tableofcontents

\section*{Introduction}

The following is a synthesis of diverse opinions and discussions among physicists who participated in the discussion session indicated above (session leader: Savannah Thais, Hunter College).  
All participating physicists are engaged in advancing physics research through the use of AI and represent major research and educational institutions across many countries.  
Accordingly, these views reflect what is currently unfolding within real research environments where AI is actively employed.

\section{Question 1: \textit{What are the values of physics in the age of AI?}}

We are entering an era in which AI can substantially accelerate research in physics, as reflected in the rapid growth of machine-learning applications across high-energy physics, lattice field theory, quantum many-body physics, and physics-informed modelling.\cite{feickert2021livingreview,boyda2022applications,carleo2017neuralquantum,karniadakis2021physicsinformed}  
Does this imply that physicists will be replaced by AI?  
Discussions were held concerning where the value of physics lies in such a time.

\subsection*{A. Predictability and Reproducibility}

Physics is a scheme in which predictions and their reproducibility hold universally, independent of who makes the claims, regardless of nationality or cultural background.  
Any proposed prediction is confirmed through experiment in nature, or, if it is mathematical in nature, by proof carried out by someone else or by an AI system.  
Even when predictions or intermediate calculations are incorrect, the community acknowledges the errors, corrects them logically or through new experiments, and thereby expands physics in a way accepted by all.  
Importantly, the highest value is ascribed to actual predictions --- i.e. statements made about observations in data before the experiment was conducted --- as opposed to post-hoc interpretations of data.
Such predictability and reproducibility constitute a shared foundation of physics for all humanity.

\textbf{Keywords:} make predictions; testability / verifiability; error correction.

\subsection*{B. Consistency With Reality}

Closely related to the above, physics advances by continually checking itself against nature, thereby uncovering truths about the universe.  
Consistency with reality is common to all natural sciences represented by physics, and objective truth---within the limits of experiment---is the driving force of scientific progress.  
Because experiments have various limitations, the natural phenomena accessible to physics are themselves bounded by such constraints.  
Thus, phenomena targeted by AI and phenomena targeted by physics may be separated by differing requirements for consistency with empirical reality. 

State-of-the-Art AI is still not exactly providing concrete technological or engineering-level upgrades to vital physics experiments; additionally, it remains unclear whether future AI growth will fill this gap.
However, machine learning has been employed in experiments for many years, and 
for instance accelerator control with reinforcement learning, Bayesian optimisation, or AI-assisted experiment design are in progress.\cite{kaiser2024accelerator,dorigo2023experimentdesign} This means that at some level AI can aid in engineering-level upgrades. 
Here the important thing is that this must always be co-creation, humans still have to fund and build machines and should have an irrevocable stake in deciding the future of experiments.

\textbf{Keywords:} true to reality; objective truth (up to limits).

\subsection*{C. Theoretical Structure}

Physics aims to uncover theoretical relations and structures behind diverse phenomena, exemplified by Occam's razor.\cite{baker2016simplicity}  
As seen in particle-physics-style reductionism, physics seeks laws and theories written with a minimal number of parameters that nonetheless explain a wide variety of phenomena.  
Emergent phenomena, as in condensed matter physics, are also within its scope; the goal is to achieve universal understanding through minimal-parameter laws, even when reductionism alone cannot explain the diversity of emergent layers.  
The aim of physics is to systematize the vast diversity of natural and numerical phenomena into laws comprehensible to humans.  
As long as this remains the goal of physics, it should naturally become clear how AI will be used within it and how it will be combined and integrated with physics. In fact, as discussed in the next paragraph on the ``black box problem," currently AI tools can accelerate physics particularly when the interpretability is not the main purpose of the problem. 
%As long as this remains the objective, the way AI is employed, combined with physics, and integrated into the discipline will naturally become clear.

\textbf{Keywords:} reductionism (to human level); emergence; systems study.

\subsection*{D. The Black Box Problem}

AI is often criticized for its ``black box'' nature, whereas physics traditionally rejects black boxes.  
This perceived dichotomy is deeply rooted in the thinking of physicists.  
However, another perspective is essential.  
In fields where AI has already significantly advanced progress---such as lattice field theory or solving Schrödinger equations in condensed matter physics---the grand challenge is not obtaining analytic expressions for ground-state wave functions, but instead computing numerical solutions efficiently.\cite{boyda2022applications,carleo2017neuralquantum}  
Thus, research progresses while accepting a certain degree of black-box methodology.  
Consequently, the extent to which interpretability must be pursued varies by subfield, and AI can assist physics at many such levels.

\textbf{Keywords:} no black boxes; which ones are acceptable?; phenomena obfuscation?

\subsection*{E. Curiosity and Uncertainty}

Research achievements celebrated in physics are not those involving mere algebraic manipulations or routine experimental refinements.  
Rather, they are those that stimulate scientific curiosity, interest, and depth among researchers.  
The importance of research is not measured by journal prestige or publication count.  
Nevertheless, there is a widespread societal inclination to evaluate physics---which spans highly diverse topics---through metrics reducible to numbers.  
Physicists have long voiced strong concern about such tendencies, and the advent of AI, which is transforming both the writing and publishing of papers, makes such concerns all the more pressing.

In the AI research community, benchmark tests are widely celebrated, including graduate-level science benchmarks used to compare frontier models with expert human performance.\cite{rein2024gpqa,openai2024o1,wang2026frontierscience}  
But will models that excel at benchmarks genuinely advance physics?  
Here lies a fundamental misunderstanding: physics is not a discipline that merely solves existing problems; it is a discipline that generates interesting questions.  
As Thomas Kuhn described, scientific activity governed by established paradigms and puzzle-solving traditions constitutes ``normal science.''\cite{kuhn1962structure}  
In contrast, ``revolutionary'' or ``anomalous'' science is driven by discoveries of new natural phenomena through technological innovation, gradually forming new grand challenges that propel science forward.  
AI may greatly accelerate normal science, but the essence of physics lies in anomalous science.  
The collective curiosity-driven process through which physicists create new physics cannot currently be replaced by AI.  
We must view the value of physics through the sociological nature of the physics community.

\textbf{Keywords:} curiosity / variation; interesting; uncertainty quantification / significance.

\subsection*{F. Solvability of Physics}

Discussion also touched upon various claims that \textit{Physics would be ``solved'' in N years}.
This is an ill-defined claim: Physics is a process that makes continually new testable predictions of nature and does not stop when a specific observation has been explained. As such, it cannot be solved.

While parts of the current activities of physicists might in the future be augmented by AI tools, this does not ``solve'' physics; similar to how the pocket calculator did not ``solve'' mathematics.
How impossibly difficult to achieve the complete ``solution of physics" could be rephrased as the difficulty in getting all the data in the universe.

It also misses another key aspect of physics: By-and-large, current progress in physics is not bounded by new ideas or theoretical predictions, but by designing, building, and running the necessary large-scale experiments or observation campaigns to test these ideas.

However, the prevalence of such public statements aimed at fundamental physics shows the important frontier role this field still carries in the public imagination (and accordingly in the aspirations of corporations).

\textbf{Keywords:} solved in three years.

\section{Question 2: \textit{What is the purpose of a PhD?}}

State-of-the-art AI models capable of high-level reasoning now possess wide-ranging knowledge and computational abilities which are increasingly being compared to those of doctoral students on difficult science benchmarks.\cite{rein2024gpqa,openai2024o1,wang2026frontierscience}  
AI may participate in research discussions, data analysis, application of theoretical frameworks, algebraic manipulation, and more.  
Does this imply that PhD students can be replaced by AI?  
If placed in the position of the PhD student, why should one train in tasks that AI can already do?  
This becomes a major challenge in motivating undergraduates and in cultivating the next generation of researchers, and also in ensuring to have funding and structure available to allow those who are motivated the chance to pursue physics.

\subsection*{A. Understanding Physics}

Students entering graduate school do so because they wish to understand the world through physics with their own minds.  
Even if AI can perform research analyses faster, students themselves have not acquired such understanding unless they do the work personally.  
The purpose of the PhD program is to support the student's process of understanding physics.  
AI entering this environment does not fundamentally change this purpose.  

However, some students may feel that tracing scientific developments already ``understood'' by AI is meaningless.  
They feel threatened. The point is that one of the reasons for doing a physics PhD is to understand physics, and AI cannot replace this fundamental purpose, and thus sometimes it is better to do things by hand even when it could be automated. i.e. the purpose of a PhD is not just to accelerate the PI's research but also to educate the student.
We must return to and reexamine the foundations of education.

\textbf{Keywords:} physics; understanding.

\subsection*{B. The PhD Program as Training for the Study of Complex Systems}

Physics is often viewed as the study of nature---for example, nuclear physics as the study of atomic nuclei.  
However, physics can also be viewed as learning methods to understand a far broader range of complex systems.  
Phenomena across scales can follow the same equations; thus physics becomes a discipline for studying systems underlying complex behavior.  
In fact, AI itself can be viewed as such a complex system, which partly explains the success of applying physics ideas to better understand AI. Moreover, the utility of these endeavors is further underscored by the significant trend for AI companies to hire individuals with physics PhDs.
PhD holders in physics who continue into non-academic careers possess the ability to construct system-level models of highly complex social phenomena.  
In this sense, the physics PhD is not merely training in natural phenomena, but training to understand and construct the systems beneath them.  
Physics, by addressing extreme phenomena, provides a vast theoretical framework through which diverse systems can be learned.  
Its utility to society is deeply tied to its essential foundation in repeatedly testable natural phenomena. Furthermore, this can help produce researchers who study AI from different viewpoint as yet another complex system.
Emphasizing these aspects within graduate education may help maintain motivation among PhD students in the AI era.

\textbf{Keywords:} how to study complex systems; training in the practice of physics; is there still a science?

\subsection*{C. Growth as a Researcher}

The PhD program is the period during which students gradually grow into independent researchers.  
If this process is neglected, researchers will not develop, and physics may ultimately collapse.  
Therefore, it is essential to enrich the PhD environment so that young researchers gain confidence, experience the joy of discovery with their own intellect, and feel a rightful sense of belonging to the community of physicists uncovering truths about the universe.  
For this reason, laboratories must explore methods for using AI and for interacting with AI in a way that supports such growth.

\textbf{Keywords:} opportunity to grow; expansion of the work students can do.

\subsection*{D. Social Value}

The salary of PhD students reflects society's expectations of universities.  
Although faculty may hesitate to view graduate students as labor hired by the university to perform specific tasks, all labor is tied to financial compensation and thus to a social contract.  
Both students and faculty are workers, and must consider how physicists and graduate students are perceived by society.  
As AI increasingly transforms society, the kinds of individuals needed in physics must be reconsidered from multiple perspectives.  
One concern is the conflation of science and technology into a single term, ``science and technology.''  
If science is pursued solely for social and/or economic utility, it loses much of its value.  
Physics exists because of human curiosity, and only incidentally do the resulting laws governing the world become technologically useful for humanity.  
Thus, immediate practical benefit is not the true value of scientific labor. As a historic example, electricity, in its current form, would not exist if not for generations of physicists' collective curiosity and incremental progress.

\textbf{Keywords:} skill transfer to economically valuable labor; societal return (contract).

\section{Question 3: \textit{What governing mechanisms for AI do we need?}}

\subsection*{A. International Competitive Structures}

As seen in the United States Genesis Mission, AI development competition in science has become severe and can inflame geopolitical tensions.\cite{whitehouse2025genesis}  
Physics is embedded within such international structures, and history---as in the Manhattan Project---shows that this can produce tragedies for humanity.  
We must continue discussing the role AI should play in the genuine advancement of physics and science.

\textbf{Keywords:} international arms race?

\subsection*{B. Equalizing Research Environments Through AI}

At present, researchers worldwide are divided between those who can access frontier AI models and those who cannot.  
The same is true for undergraduates and graduate students.  
This raises the question of how we should regard a world in which only those with access to AI accelerate their research while others are left behind.  
An analogy may be found in the late twentieth century, when the emergence of personal computers drastically transformed research environments.\cite{ceruzzi2003history}

\textbf{Keywords:} AI as playing-field leveler?

\subsection*{C. Research Quality and Publication Standards}

A pressing concern is that AI may mass-produce papers, altering the value of each publication and overwhelming human capacity for reading.  
Peer review systems may collapse.  
If AI performs peer review, the diversity of science may be threatened at its core.  
One simple solution would be: humans perform all peer review; AI performs only logical verification; research that AI can conduct independently is not published.  
However, given that many researchers already use AI in various aspects of research, excluding all AI-performable research from peer review may be overly simplistic.
Furthermore, AI-performable research may still contain useful insights which are beneficial to the community. 
Thus, a nuanced policy for AI's role in research is required.

A concrete example of AI's benefit is its ability to translate languages. Language differences have long been major obstacles for non-English-speaking researchers.  
Advances in neural machine translation and large language models appear to reduce this barrier.\cite{wu2016gnmt,vaswani2017attention}  
AI acceleration of research thus also has important benefits.

ArXiv's computer-science category has already adopted an updated moderation practice under which review/survey articles and position papers generally require acceptance at a peer-reviewed venue before submission, explicitly noting the influx of LLM-generated review-style manuscripts.\cite{arxiv2025reviewpolicy}  
Publication policies must adapt flexibly to the capabilities of AI.  
It may also be possible to maintain journals that require AI-based review and journals that require human-only review, treating human-understood physics and AI-driven physics as distinct scientific modes.  
Just as ``computational science'' emerged as a new scientific paradigm, we are entering an era in which new forms of science must be explored.\cite{denning2005computing,hey2009fourthparadigm}  
Is AI science merely an expansion of computational science, or is it a new engineering-like field directed toward technological application without understanding laws?  
We must carefully monitor the direction of science.  
Based on that, we must maintain, revise, or flexibly adapt the core principles of publication and peer review.

\textbf{Keywords:} higher bar for publication (if AI can do it $\rightarrow$ no paper).

\section{Prospect}

The above is a synthesis of diverse perspectives exchanged during the discussion session, organized around the keywords raised.
These do not constitute definitive answers; answers must be debated within various communities of physics and the cultural values that shape them.  
We sincerely hope that the perspectives presented here contribute to such future discussions.

\bibliographystyle{unsrt}
\bibliography{references}

\end{document}